# Evidence of partial gap opening and Ce-site dilution effects in a heavy fermion compound CeNiGe$_2$

Karan Singh and K. Mukherjee*

School of Basic Sciences, Indian Institute of Technology Mandi, Mandi 175005, Himachal Pradesh, India

*Email: kaustav@iitmandi.ac.in

## ABSTRACT

We report the results of magnetization, non-linear dc susceptibility, electrical transport, and heat capacity measurements on Y-substituted heavy fermion CeNiGe$_2$. Investigations are carried out on the compounds CeNiGe$_2$, Ce$_{0.9}$Y$_{0.1}$NiGe$_2$, Ce$_{0.8}$Y$_{0.2}$NiGe$_2$ and Ce$_{0.6}$Y$_{0.4}$NiGe$_2$. It is observed that with the increase in Y-concentration, the magnetic ordering temperature decreases. For CeNiGe$_2$, below ordering temperature Arrott plots suggest the presence of spin density wave (SDW). Third and fifth order dc susceptibility indicates magnetic instability which possibly leads to partial gap opening resulting in the observation of SDW. These observations are further investigated through resistivity and heat capacity measurements which also point toward partial gap opening in CeNiGe$_2$. Interestingly, with the increase in Y-substitution, it is noted that the gap opening is suppressed and also shifted towards lower temperature. Moreover, our investigations reveal absence of non-Fermi liquid behavior or zero field quantum critical point even after 40% dilution of Ce-site.

# 1. Introduction

The hybridization between $4f$ and conducting electrons results in fascinating ground state due to the competition between inter-site Ruderman-Kittel-Kasuya-Yosida (RKKY), and intra-site Kondo interaction in Ce-based intermetallic compounds [1, 2]. In such systems, external parameters like magnetic field, external pressure and composition is used to tune the ground state, leading to intriguing phases, such magnetic instability, non-Fermi liquid behavior etc. [3-8]. Also literature reports in some heavy fermion systems suggest the phenomenon of spin density wave (SDW) owing to the presence of a partially gapping of Fermi surface [9, 10]. The heavy fermion compound $CeNiGe_2$ has been investigated in the last decade due to the intriguing physics exhibited by this compound [11-15]. Recent results indicate that external pressure and high magnetic field are incapable of repressing the magnetic ordering, resulting in the absence of Quantum critical point (QCP) and the behavior of this compound is not in accordance to the Doniach model, which is widely used to classify heavy fermion compounds [16, 17]. Kim et al. reported that partial replacement of Ge by Si results in suppression of magnetic ordering and some signatures of Quantum critical point (QCP) is observed [18]. However to the best of our knowledge there is no literature reports about QCP or signature of non-fermi liquid behavior observed due to Ce-site dilution by a nonmagnetic ion in $CeNiGe_2$ compound. Also, the low temperature ground state of $CeNiGe_2$ which has received considerable attention in the past decade remains undetermined. In these heavy fermions systems an enhanced magnetic susceptibility is a natural outcome, providing the possibility of measureable large non-linear magnetic susceptibility, which is an important tool to probe exotic type of spin ordering which in turn is helpful in determining the ground state of a compound.

Hence, in this manuscript, we report the results of magnetization, non-linear dc susceptibility, transport and thermodynamic measurements of Y-substituted $CeNiGe_2$. Studies are carried out on the following compounds: $CeNiGe_2$, $Ce_{0.9}Y_{0.1}NiGe_2$, $Ce_{0.8}Y_{0.2}NiGe_2$ and $Ce_{0.6}Y_{0.4}NiGe_2$. The parent compound reported undergoes two antiferromagnetic orderings around 4.1 and 3 K [11, 12, 17]. The fundamental observations are: i) with Y-substitution the two transition temperature decreases for $Ce_{0.9}Y_{0.1}NiGe_2$ while a single transition is observed for $Ce_{0.8}Y_{0.2}NiGe_2$, ii) for $Ce_{0.6}Y_{0.4}NiGe_2$ compound the transition is suppressed below 1.8 K, iii) Arrott plots suggest the presence of SDW iv) non-linear susceptibility indicates magnetic instability which possibly leads to partial gap opening resulting in the observation of SDW v) zero field resistivity and heat capacity results also indicates to the presence of partially gap vi) with the increase in Y-substitution the gap opening is suppressed and also shifted down in temperature and vii) magnetic Gruneisen parameter gives an indication of presence of unusual fluctuations for Y-0.4 compound.

## 2. Experimental details

The polycrystalline compounds $Ce_{0.9}Y_{0.1}NiGe_2$ (Y-0.1), $Ce_{0.8}Y_{0.2}NiGe_2$ (Y-0.2) and $Ce_{0.6}Y_{0.4}NiGe_2$ (Y-0.4) are prepared by arc melting and the stoichiometric amounts of respective high purity elements under similar conditions as reported in Ref [17]. The $CeNiGe_2$ (Y-0.0) compound is the same as used in Ref [17]. The characterization of compounds is done by x-ray diffraction (XRD) studies at room temperature which established single phase nature of these compounds (within the detection limit of the technique) (Fig. 1). The XRD pattern is indexed to orthorhombic type $CeNiSi_2$-type structure. It is also observed that with increase in Y-substitution peak intensity shift to higher angle (inset of Fig. 1), thereby establishing that the dopant go to respective site (the curve for Y-0.0 compound is added from ref [17] for comparison). We also performed the energy dispersive spectroscopy (EDAX) to get confirmation about the average atomic stoichiometry, which is found to be in accordance with the expected values. Temperature (*T*) dependent magnetization (*M*) are performed using Magnetic Property Measurement System (MPMS), while, temperature and magnetic field (*H*) dependent heat capacity (*C*) and resistivity (ρ) measurements are performed using Physical Property Measurement System (PPMS), both from Quantum design, USA. *M*, *C* and ρ measurements were carried out on pellets of specific shapes.

## 3. Results and discussions

Fig. 2 (a) shows the temperature response dc susceptibility (*M/H*) under zero field cooling condition for all the compounds at 0.005 T upto 20 K. The two antiferromagnetic transition temperatures at ~ 4.1 ($T^I_N$) and 3 K ($T^{II}_N$) observed for Y-0.0 compound is in accordance to that reported in literature for single crystals [11, 12]. This transition temperature is shifted down to ~ 3 and 2.2 K for Y-0.1 compound. In Y-0.2 compound, a single peak is noted and the compound undergoes a transition at ~ 2 K. For the extreme composition i.e. Y-0.4 compound, the transition temperature is suppressed below 1.8 K (Fig. 2(b)). The observation of decreasing transition temperature with increasing Y-concentration indicates that exchange interaction among spins suppress significantly, implying, Ce-site dilution effect. In all these compounds no bifurcation is observed between the magnetization curves obtained under zero field cooled and field cooled condition (not shown) which point to the fact that magnetocrystalline anisotropy plays an insignificant role in these polycrystalline compounds of the series $Ce_{1-x}Y_xNiGe_2$ [19]. Inverse magnetic susceptibility of the Y-substituted compound is fitted with Curie Weiss law in the temperature range of 100 – 300 K (Fig. 2(c)), as below 100 K, a non-linear deviation is observed due to crystalline electric field (CEF) effect [11, 17]. The obtained effective moment ($\mu_{eff}$) and the Curie Weiss temperature (θ) is plotted in Figs. 2 (d) and (e) respectively. The value of these parameters for Y-0.0 compound is added from Ref [17] for comparison. The theoretical effective magnetic moment $\mu_{eff}$, is calculated using the expression: $\mu_{eff} = xg [(J(J+1)]^{1/2}$, where $x = 1, 0.9, 0.8,$ and $0.6$ for Y-0.0, Y-0.1, Y-0.2, and Y-0.4 compound respectively. $g$ is Lande's splitting factor and $J$ is the total angular momentum. For $Ce^{3+}$, the $g$ and $J$ are 6/7 and 5/2 respectively. The calculated $\mu_{eff}$ is shown in Fig. 2 (d). It is noted that $\mu_{eff}$ decreases with increase of Y-substitution. Also, as observed from the Fig. 1 (d), the obtained experimental value matches well with the calculated theoretical values, except for Y-0.4 compound which may be ascribed to lattice contraction effects on magnetic interaction. It is noted that θ shows an unsystematic variation with Y-substitution, suggesting that the internal chemical pressure alter the exchange interaction from the dominating antiferromagnetic in Y-0.0 and Y-0.1 compounds to ferromagnetic at Y-0.2 compound. However, for Y-0.4 compound, it again indicates the dominance antiferromagnetic interactions. Such type of unsystematic variation in θ has also been noted for other compound series [20]. To get an insight about the magnetic state of these compounds, isothermal magnetization $M (H)$ data at 2 K are obtained (Fig. 2 (f)). The metamagnetic transition at ~ 0.7 T observed in Y-0.0 compound [11, 12, 17] is suppressed with Y-substitution. In order to have an idea about the delocalization of magnetic moments in this series, we have calculated the ratio of $\mu_{eff}/\mu_s$ where, $\mu_{eff}$ is the number of magnetic carriers which is deduced from Curie Weiss law. The value of $\mu_s$ is deduced from the magnetic moment at 2 K at 7 T. According Takahashi plot (which is modified form of Rhodes-Wohlfarth plot [21, 22]), when the ratio of $\mu_{eff}/\mu_s \sim 1$ it indicates that moments are localized on lattice sites, while $\mu_{eff}/\mu_s > 1$ point towards deviation from localized to itinerant moment [21]. The value of $\mu_{eff}/\mu_s$ is obtained from the expression of curie constant: $C = (\mu_B^2/3R)\mu^2_{eff}$; where $\mu_{eff}$ is the effective moment of carriers. The estimated values of $\mu_{eff}/\mu_s$ are ~ 2.8, 3.0, 3.2, and 3.8 for Y-0.0, Y-0.1,

Y-0.2, and Y-0.4 compounds respectively. From this observation we infer that all compounds shows the existence of itinerant moment, which is found to be increase with Y-substitution.

In order to further study the magnetic state of these compounds, Arrott plots are done. Fig. 3 (a, b, c, and d) shows $M^2$ versus $H/M$ plots for all the compounds at different temperatures. For Y-0.0 compound, it is observed that the curves are linear at high field and it changes in the form of an arc with negative slope below 3.9 K (inset of Fig. 2 (a)). According to the criteria suggested by Banerjee [23], the negative and positive slope in Arrott plots indicate to the first and second order of transition respectively. Also for first order transition the value of the negative slope should increase with increasing temperature which is in contrast to that observed for Y-0.0 compound where the value of the negative slope decreases. Such opposite behavior has been studied by taking into account the asymmetry of the free energy density with respect to the magnetization and it was observed that a negative slope is possible at second order phase transition with the slope value decreasing with increasing temperature [24]. Also the first order transition generally results in hysteresis in *M-H* curve at metamagnetic transition field $H_c$. Even though the Y-0.0 compound exhibits the metamagnetic transition at $H_c$ = 0.7 T (below the ordering temperature of the compound) [17], the *M-H* curve does not shows hysteresis at $H_c$. The observed feature in the Arrott plots in this compound has also been reported in compounds like $Nb_{1-y}Fe_{2+y}$, where the feature was ascribed to SDW behaviour [25]. Therefore, it can be said that Y-0.0 compound shows signature of SDW below ordering temperature. This slope change of Arrott plots goes down to ~ 3 and 1.8 K respectively for Y-0.1 and Y-0.2 compounds. For Y-0.4, no such changes in slope appear in the measured temperature range. This indicates that the temperature for the onset of SDW gets shifted down in temperature with Y-substitution.

Non-linear dc magnetic susceptibility is an important tool to probe the nature of exotic spin ordering in magnetic ground state of different compounds. The non-linear parts of dc susceptibilities are obtained from field and temperature dependence of magnetization. The non-linear susceptibilities are expressed as, $M = \chi_1 H + \chi_3 H^3 + \chi_5 H^5..$, where $\chi_1$ is the linear magnetic susceptibility, $\chi_3$, $\chi_5$,…. are the non-linear magnetic susceptibility. The data for this measurement are collected under the field cooled condition in the field range of 0 – 1 T by the protocol described in Ref [26]. The compound is first cooled from 300 K to 20 K in zero field and then 50 Oe magnetic field was switched on. Under field the compound is cooled down to temperature 1.8 K at the rate of 5 mK/sec. Then the temperature response of the magnetization is measured in the range of 1.8 to 20 K. At 20 K, the next field is applied and the same protocol is repeated. This procedure is generally followed as it diminishes the effect of magnetic hysteresis of the superconducting magnet of MPMS on the measured data. Thus, by plotting the ratio $M/H$ versus $H^2$ (Fig 4: (a), (b), (c), and (d)) at different temperature, $\chi_3$ (coefficient of $H^2$) and $\chi_5$ (coefficient of $H^4$) are separated from a quadratic fits of the above equation to the experimental data at a particular temperature, while $\chi_1$ is obtained from the intercept. Temperature response of this $\chi_1$ is plotted as shown in the inset of Fig. 5 (a). The nature of curves for all the compounds replicate to that observed from dc magnetization measurements. Fig. 5 (a) represents the temperature response of this $\chi_3$. For Y-0.0 compound it is observed that above ~ 4.1 K, $\chi_3$ is negative while it changes its sign below this temperature. Interestingly two peaks are observed ~ 3.8 and 3.2 K which is near the transition temperature of this compound. Chandra et al. had reported that $\chi_3$ can be used to probe the presence of quadrupolar moments [27]. Hence, the observation of finite $\chi_3$ suggests the presence of quadrupolar moment in Y-0.0 compound. This indicates to the presence of higher order magnetization, which signals to the existence of exotic magnetic state in this compound. Here it is to be noted that, even though a doublet ground state is expected for $Ce^{3+}$ ions, some Ce based compounds like $CeRu_2Si_2$ and $Ce_{1-x}Y_xRu_2Si_2$ exhibit finite $\chi_3$ which is related to quadrupolar moment [28]. The magnitude of $\chi_3$ is also seen to decrease with the increase of Y-substitution. For Y-0.1 compound $\chi_3$ is positive below ~ 3.1 K while for Y-0.2 compound the crossover temperature shift to ~ 2.0 K. For Y-0.4 compound, $\chi_3$ is negative in the temperature of measurement. The temperature response of $\chi_5$ is displayed in inset of Fig. 5 (b). For Y-0.0 compound, it is observed that $\chi_5$ is positive below ~ 3.8 K with the presence of a weak anomaly around second transition temperature. This anomaly shifts down to ~ 3.1 K and 2.0 K for Y-0.1 and Y-0.2 compounds respectively. For Y-0.4 compounds, $\chi_5$ is negative in temperature range of measurement. We further investigated the non-linear susceptibility with a general equation of magnetic free energy ($F$) [29]; $F = - HM + a_2M^2 + a_4M^4 + a_6M^6 + ....$, where $a_2 = 1/2\chi_1$, $a_4 = - \chi_3/4\chi_1$, and $a_6 = [3\chi^2_3 - \chi_5\chi_1]/\chi^7_1$ are the coefficients of expanded $F$. The quantity $(3\chi^2_3 - \chi_5\chi_1)$ is a measure of stability of the magnetic state. Temperature variation of this quantity is shown in Fig. 5 (b). As seen from the Fig., it is negligible above ~ 4.0 K, while a positive response is observed within temperature range of ~ 3.6-4.0 K. Again a crossover is observed below ~ 3.6 K even though $\chi_3$ and $\chi_5$ are positive. This observed crossover can be ascribed to instability of magnetic state [29] which might be open the partial gap below ~ 3.6 K. For Y-0.1 compound the crossover temperature shift to 3.0 K. For Y-0.2, and Y-0.4 compounds it get suppressed below 1.8 K. This reveals that with Y-substitution the gap is being suppressed. The observed gap opening can be related to onset of SDW which to stabilize the magnetic state in Y-0.0 compound. This feature is not uncommon and has also been reported in the SDW region of other heavy fermion compounds [27, 30].

Fig. 5 (c) shows the temperature dependence of normalized electrical resistivity $\rho(T)/\rho_{50K}$ at zero field for all compounds in the temperature range 1.8 to 30 K. For Y-0.0 compound a maxima in resistivity is seen to around 5 K and on decreasing temperature resistivity decreases followed by a slope change near 3.2 K (inset of figure 5 (c)). Below 3.2 K, the resistivity was found to vary linearly. Hence the observed linear dependence in resistivity in this compound is ascribed to spin disorder which induces scattering of carriers [11, 12]. Also it is expected that presence of a gap results in carrier losses, hence resistivity is expected to increase. However for Y-0.0 it was found the resistivity decreases as temperature is reduced below the gap. The observed decrease in resistivity is probably due to the reduction of the scattering rate which overcompensate the depletion of the carriers [31]. With Y-substitution the gap opening is suppressed below 1.8 K. Hence the decrease in resistivity with reduction in temperature is not observed in these compounds, and it is found that resistivity increases with Y-substitution, which might be due to increasing spin disorder and residual resistivity.

To give further indirect evidence about the partial gap opening, heat capacity of these compounds is measured. Fig. 6 (a) shows the temperature response of *C/T*. The observed transitions temperatures are ~ 3.9 and 3.1 K for Y-0.0 compound. It is shifted down to ~ 3.1 K (weak anomaly) and 2.2 K for Y-0.1 compound and ~ 2.2 K for Y-0.2 compound. For Y-0.4 compound the transition is suppressed below 1.8 K. The observed transition temperatures of these compounds are in analogy to that obtained from magnetic measurements. Inset of Fig 6 (b) shows the temperature response of 4*f*-electron contribution to heat capacity ($C_{4f}$) obtained from Ref [17]. It is observed that behavior of $C_{4f}$ varies exponentially below transition temperatures. Hence we have a small temperature range to do the fitting for Y-0.0 compound and also for other Y-doped compounds. Therefore we analysed our data according to standard BCS expression ($\Delta C_{4f}/\gamma T_N = 1.43$) where the formation of SDW associated with gap opening and is also consistent with a jump in heat capacity at transition temperatures [32, 33]. In Y-0.0, the jump in the $C_{4f}$ is observed at transition temperature $T^I_N$ and using relation: $\Delta C_{4f} = C_{4f}(T^I_N) - C_{4f}(n)$, (where $C_{4f}$ ($T^I_N$) is the heat capacity at $T^I_N$ and $C_{4f}$ ($n$) is heat capacity in the paramagnetic region), is found to be 2.43 J/mol-K$^2$. The value of γ obtained from Kondo temperature ($T_K$) was 0.433 J/mol-K$^2$ [17]. The value of the factor $\Delta C_{4f}/\gamma T_N$ was found to be around 1.4, which is consistent with the value of standard BCS expression and SDW hypothesis. Similar type of analysis has been carried in other compounds which show SDW behavior [34, 35]. From this observation, we can say that SDW is associated with gap opening in Y-0.0 compound. Such observation of SDW is not uncommon and has been reported in other Ce-based polycrystalline compound [33]. It is also observed that with the increase in magnetic field this gap is suppressed, however, a new antiferromagnetic state is noted as is reported in Ref [17].

We also studies the temperature dependence magnetic entropy ($S_{4f}$), (calculated using the similar procedure as observed in Ref [17]) as shown in Fig. 6 (b). For Y-0.4 compound, we did not calculate $S_{4f}$, as the magnetic state for this compound gets suppressed below 1.8 K. From the Fig. 6 (b) it is noted that with increasing Y-concentration, the entropy decreases above the ordering temperature due to increased screening of magnetic moment by conducting electrons. For Y-0.1 compound it is observed that below the ordering temperature (shown by arrow) entropy increases as compared to Y-0.0 compound. As for the Y-0.1 compound the gap is suppressed and shifted down in temperature which results in increase in entropy with the increase in accessible state [34]. Thus our result indicates that with Y-substitution the gap is suppressed which is also in analogy with the results of the previous sections. However, such effects are not visible for Y-0.2 and Y-0.4 compounds as the magnetic transition in these compounds are suppressed to a very low temperature.

Additionally, the temperature dependence of magnetic Gruneisen parameter ($\Gamma_{mag}$) at 0.1 T is studied. The $\Gamma_{mag}$ is the sensitive tool to estimate the degree of freedom of delocalized spins, and categorize quantum critical point (QCP) [36]. It is calculated using the equation [37], $\Gamma_{mag}$ = -d$M$/d$T$*$C$; where d$M$/d$T$ is the temperature derivative of magnetization under zero field cooled (ZFC) condition at 0.1 T. Fig. 6 (c) shows the temperature response of $\Gamma_{mag}$. A crossover from positive to negative value at 4.0 K is observed in Y-0.0 compound, followed by a dip at ~ 3.6 K. The sign change occurs due to accumulation of entropy at phase boundary in critical regime [37]. Above 4 K, positive Γ indicates paramagnetic regime and a crossover at ~ 4 K indicates entropy accumulation. The peak gets suppressed and the peak temperature also decreases to ~ 2.9 K for Y-0.1 compound. This feature is suppressed below 1.8 K for Y-0.2 and 0.4 compounds. Interestingly for Y-0.4 compound Γ is seem to diverge. This curve is fitted with a power law of the form [38], $\Gamma_{mag} \sim T^{\varepsilon}$; where the ε is temperature exponent. It is observed that the curve could be fitted with this equation below 2.5 K with ε ~ -2.3. This parameter is incompatible with the prediction of itinerant theory, where $\Gamma_{mag} \sim T^1$ [37]. However similar variation have been reported in compound like $YbRh_2Si_2$ and $Ce(Ni_{1-x}Pd_x)_2Ge_2$ [38, 39], which incompatible with itinerant theory. Hence it can be said that in Y-0.4 compound unusual fluctuation are present. Even though a divergence in $\Gamma_{mag}$ suggests that a system is near the QCP, however, for Y-0.4 compound no other signatures of QCP like zero field quantum critical scaling, logarithmic or square root dependence of *C* etc. are observed. Further experiments, using magnetic field or pressure *P*, would be interesting to, investigate this unusual divergence or proximity of quantum criticality in Y-0.4 compound.

## 4. Conclusion

In summary, investigation of magnetization, non-linear dc susceptibility, transport and heat capacity measurements for the Y-substituted heavy fermion system $CeNiGe_2$ is reported. Y-substitution results in sequential suppression of magnetic ordering temperature below 1.8 K and also increases itinerancy among the spins. In $CeNiGe_2$ below ordering temperature, Arrott plots suggest towards the observation of SDW. Analysis of non-linear susceptibility indicates magnetic instability which possibly leads to partial gap opening resulting in the observation of SDW. Results of resistivity and heat capacity also support these observations. This gap opening is suppressed and also is shifted down in temperature with the increase in Y-substitution. Moreover, signature of non-Fermi liquid behavior or zero field QCP is absent even after 40% dilution of Ce-site, unlike to that, observed for Si substituted $CeNiGe_2$. Studies of magnetic Gruneisen parameter give an indication of presence of unusual fluctuations in Y-0.4 compound. This work may stimulate investigation, by using direct probe methods like inelastic neutron diffraction scattering and μSr to further investigate the magnetic ground state of such heavy fermion compound series.

**Acknowledgements**

The authors acknowledge experimental facilities of Advanced Material Research Centre (AMRC), IIT Mandi. Financial support from IIT Mandi is also acknowledged.

**References**

[1] N. Grewe, and F. Steglich, *Handbook on the Physics and Chemistry of Rare Earths*, Elsevier, Amsterdam, 1991, 14, pp.343

[2] S. Doniach, *Valence Instabilities and Related Narrow Band Phenomena*, Parks Plenum, New York, 1977, pp. 169

[3] H. v. Lohneysen, A. Rosch, M. Vojta, and P. Wolfle, Rev. Mod. Phys. 79 (2007), pp. 1015-1075

[4] J. R. Jeffries, N. A. Frederick, E. D. Bauer, H. Kimura, V. S. Zapf, K. D. Hof, T. A. Sayles, and M. B. Maple, Phys. Rev. B 72 (2005), pp. 024551

[5] H. v. Lohneysen, T. Pietrus, G. Portisch, H. G. Schlager, A. Schroder, M. Sieck, and T. Trappmann, Phys. Rev. Lett. 72 (1994), pp. 3262-3265

[6] P. Gegenwart, J. Custers, C. Geibel, K. Neumaier, T. Tayama, K. Tenya, O. Trovarelli, and F. Steglich, Phys. Rev. Lett. 89 (2002), pp. 056402

[7] P. Schlottmann, in *Handbook of Magnetic Materials*, vol. 23, Chapter 2, K. H. J. Buschow, editor, Elsevier B. V. (2015), pp.85.

[8] A. Yeh, Y.-Ah Soh, J. Brooke, G. Aeppli, T. F. Rosenbaum, and S. M. Hayden, Nature 419 (2002), pp. 459-462

[9] R. Movshovich, A. Lacerda, P. C. Canfield, J. D. Thompson and Z. Fisk, Phys. Rev. Lett. 73 (1994), pp. 492-495

[10] H.-F Li, C. Cao, A. Wildes, W. Schmidt, K. Schmalzl, B. Hou, L.-P Regnault, C. Zhang, P. Meuffels, W. Loser, and G. Roth, Sci Rep. 5 (2015), pp.7968

[11] M. H. Jung, N. Harrison, A. H. Lacerda, H. Nakotte, P. G. Pagliuso, J. L. Sarrao, and J. D. Thompson, Phys. Rev. B 66 (2002), pp. 054420

[12] V. K. Pecharsky V K, and K. A. Gschneidner, Jr., Phys. Rev. B 43 (1991), pp. 10906

[13] A.P.Pikul, D. Kaczorowski, Z. Bukowski, T. Plackowski, and K. Gofryk, J. Magn. Magn. Mater. 16 (2004), pp. 6119-6128

[14] C. Geibel, C. Kammerer, B. Seidel, C.D.Bredl, A. Grauel, and F. Steglich, J. Magn. Magn. Mater. 108 (1992), pp. 207

[15] P. Schobinger-Papamantellos, A. Krimmel, A. Grauel, and K.H.J. Buschow, J. Magn. Magn. Mater. 125 (1993), pp. 151

[16] A. T. Holmes, T. Muramatsu, D. Kaczorowski, Z. Bukowski, T. Kagayama, and K. Shimizu, Phys. Rev. B 85 (2012), pp. 033101

[17] K. Singh, and K Mukherjee, Phys. Lett. A 381 (2017), pp. 3236

[18] D. Y. Kim, D. H. Ryu, J. B. Hong, J.-G Park, Y. S. Kwon, M. A. Jung, M. H. Jung, N. Takeda, M. Ishikawa, and S. Kimura, J. Phys.: Condens. Matter 16 (2004), pp. 8323

[19] P. A. Joy, and S. K. Date, J. Magn. Magn. Mater. 218 (2000), pp. 229

[20] K. Huang, J. J. Hamlin, R. E. Baumbach, M. Janoschek, N. Kanchanavatee, D. A. Zocco, F. Ronning, and M. B. Maple, Phys. Rev. B 87 (2013), pp. 054513

[21] Y. Takahashi, J. Phys. Soc. Jpn. 55, (1986), pp. 3553-3573

[22] P. Rhodes, and E. P. Wohlfarth, Proc. R. Soc. Lond. A 273 (1963), pp. 247-258

[23] S. K. Banerjee, Phys. Lett. 12 (1964), pp. 16

[24] S. Bustingorry, F. Pomiro, G. Aurelio, and J. Curiale, Phys. Rev. B 93 (2016), pp. 224429 (R)

[25] D. Moroni-Klementowicz, M. Brando, C. Albrecht, W. J. Duncan, F. M. Grosche, D. Gruner, and G. Kreiner, Phys. Rev. B 79 (2009), pp. 224410

[26] M. J. P. Gingras, C. V. Stager, N. P. Raju, B. D. Gaulin, and J. E. Greedan, Phys. Rev. Lett. 78 (1997), pp. 947

[27] P. Chandra, A. P. Ramirez, P. Coleman, E. Bruck, A. A. Menovsky, Z. Fisk, and E. Bucher, Physica B 199 (1994), pp. 426

[28] J.-G Park, P. Haen, P. Lejay, and J. Voiron, J. Phys.: Condens. Matter 6 (1994), pp. 9383-9392

[29] B. S. Shivaram, B. Dorsey, D. G. Hinks, and P. Kumar, Phys. Rev. B 89 (2014), pp. 161108 (R)

[30] A. P. Ramirez, P. Coleman, P. Chandra, E. Bruck, A. A. Menovsky, Z. Fisk, and E. Bucher, Phys. Rev. Lett. 68 (1992), pp. 2680

[31] F. Steckel, S. Rodan, R. Hermann, C. G. F. Blum, S. Wurmehl, B. Buchner, and C. Hess, Phys. Rev. B 90 (2014), pp. 134411.

[32] S. Murayama, C. Sekine, A. Yokoyanagi, K. Hoshi, and Y. Onuki, Phys. Rev. B 56 (1997), pp. 11092

[33] J. Tang, T. Matsumoto, H. Abe, and A. Matsushita, Solid state commun. 109 (1999), pp. 445

[34] J. Wooldridge, D. M. K. Paul, G. Balakrishnan, and M. R. Lees, J. Phys.: Condens. Matter 17 (2005), pp. 707

[35] H. Gu, J. Tang, and A. Matsushita, Phys. Rev. B 65 (2001), pp. 024403

[36] M. Garst, and A. Rosch, Phy. Rev. B 72 (2005), pp. 205129

[37] L. Zhu, M. Garst, A. Rosch, and Q Si, Phys. Rev. Lett. 91 (2003), pp. 066404

[38] Y. Tokiwa, T. Radu, C. Geibel, F. Steglich, and P. Gegenwart, Phy. Rev. Lett 102 (2009), pp. 066401

[39] P. Gegenwart, Y. Tokiwa, J. G. Donath, R. Kuchler, C. Bergmann, H.S. Jeevan, E. D. Bauer, J. L. Sarrao, C. Geibel, and F. Steglich, J. Low Temp Phys 161 (2010), pp. 117

**Figures**

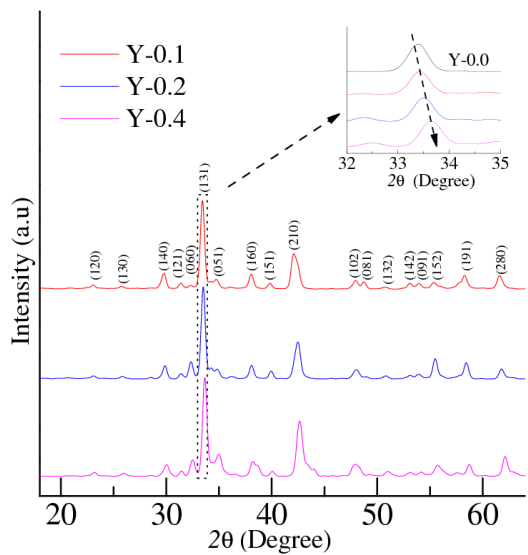

**Figure 1:** XRD patterns for Y-0.1, Y-0.2, and Y-0.4 compounds. Inset: The region around the main peaks is shown in an expanded form to show a gradual shift of diffraction lines with changing composition. The XRD pattern for Y-0.0 compound is taken from Ref [17] for comparison.

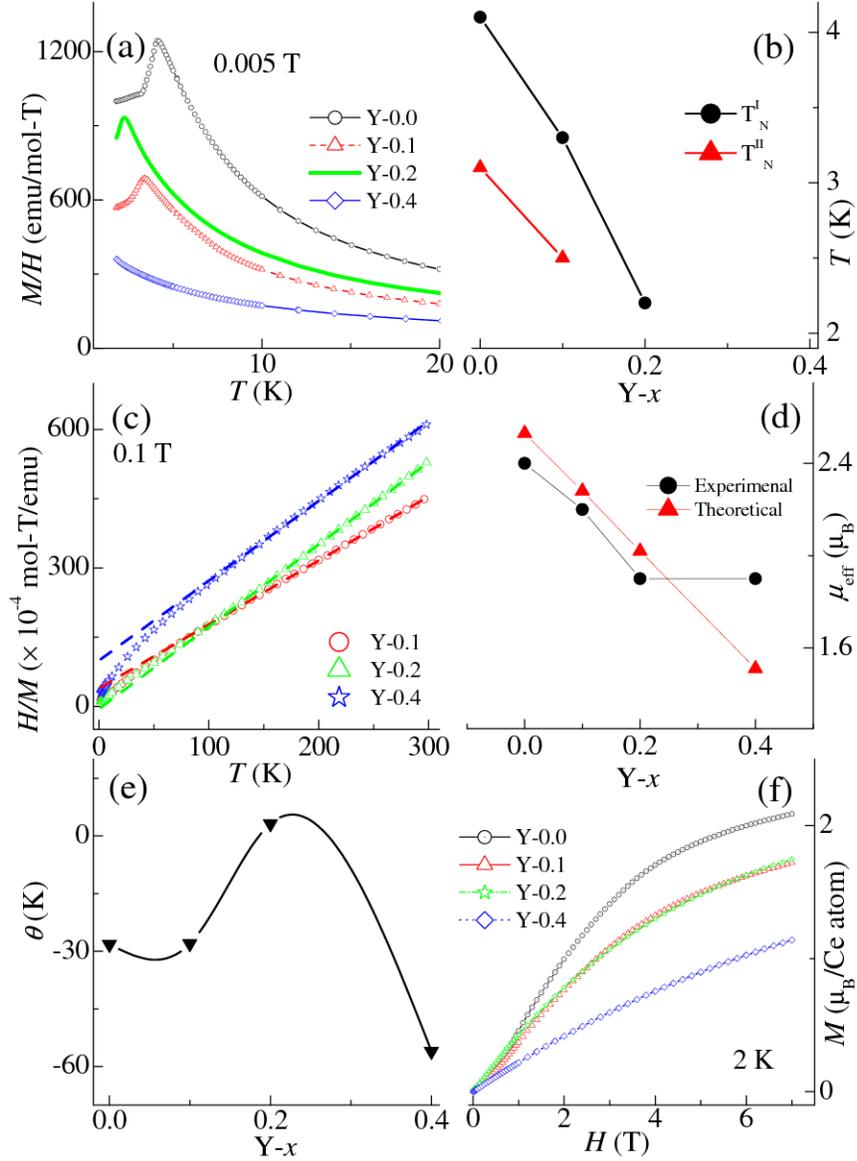

**Figure 2**: (a) Temperature response of dc susceptibility (*M/H*) under zero field cooling condition at 0.005 T for all the compounds. (b) Variation of transition temperature with Y-concentration. (c) Temperature response of inverse magnetic susceptibility (*H/M*) at 0.1 T. Straight lines through the curves shows the Curie Weiss law fitting. (d) and (e) Variation of effective moment ($\mu_{eff}$) and Curie Weiss temperature ($\theta$ (K)) with Y-concentration, respectively. (f) Isothermal magnetization versus magnetic field at temperature 2 K for all the compounds. The data for Y-0.0 is taken from Ref [17] for comparison.

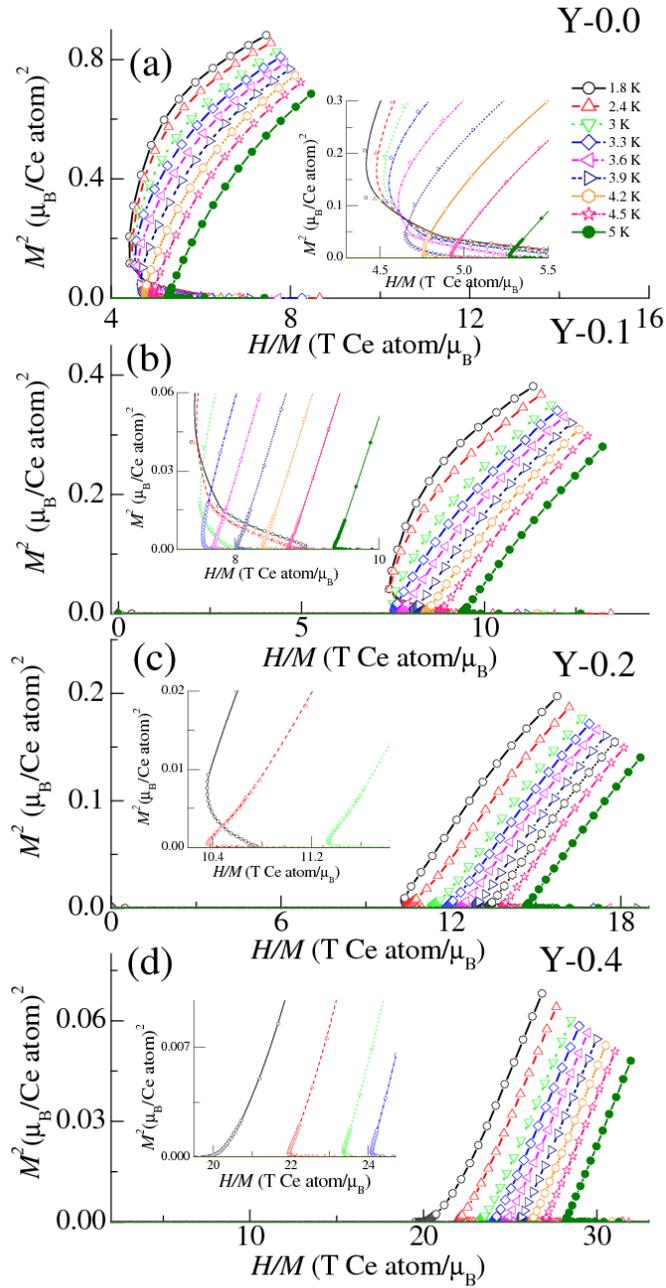

**Figure 3**: (a), (b), (c), and (d) $M^2$ versus $H/M$ for plots for all the compounds in the temperature range of 1.8 - 5K. Inset: Same plots magnified to highlight the change of slope.

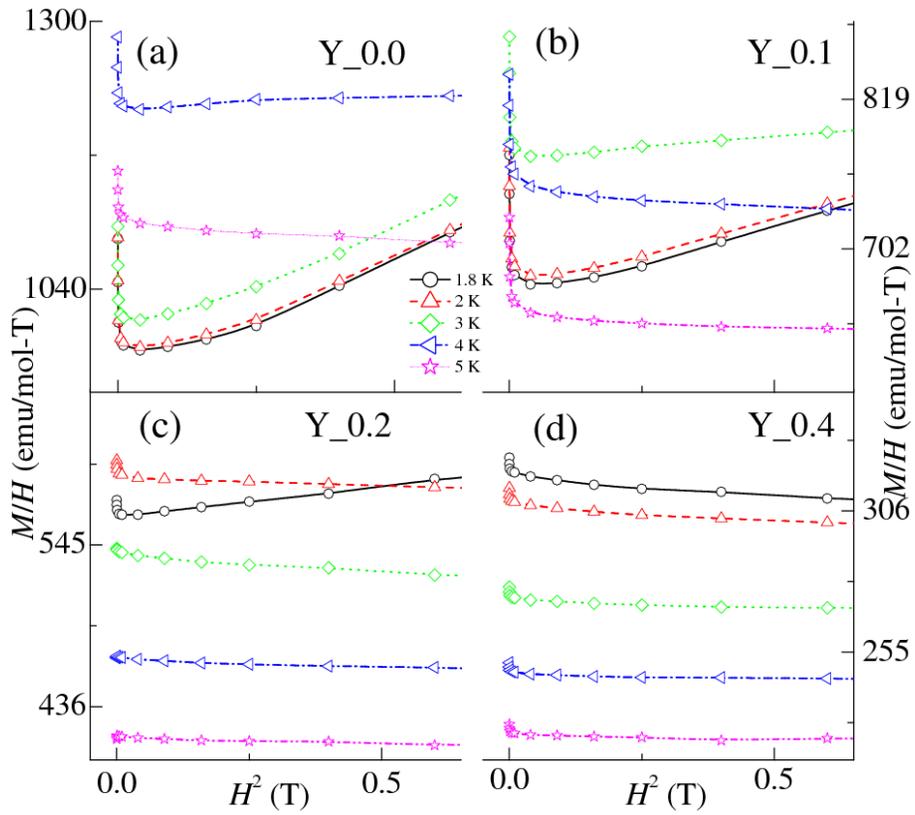

**Figure 4**: (a), (b), (c), and (d) shows the $M/H$ vers $H^2$ plots for Y-0.0, Y-0.1, Y-0.2, and Y-0.4 respectively.

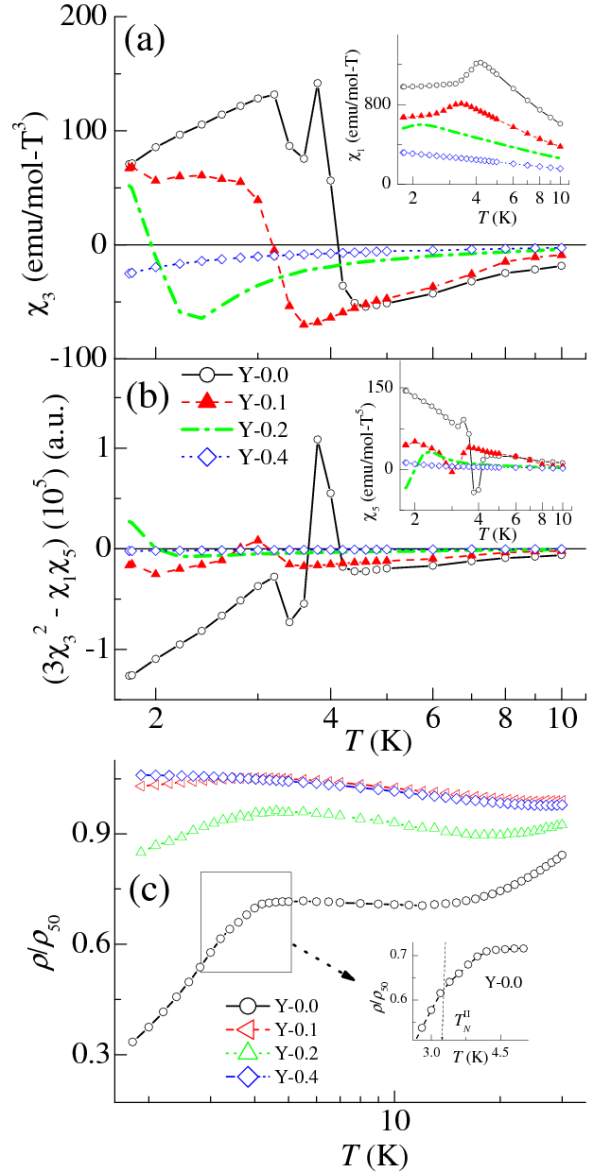

**Figure 5**: (a) and its inset: Temperature response of third order susceptibility ($\chi_3$) and first order susceptibility ($\chi_1$) for the series of compounds $Ce_{1-x}Y_xNiGe_2$ (x = 0.0, 0.1, 0.2, and 0.4) respectively. (b) and its inset: Temperature response of the quantity ($3\chi_3^2 - \chi_5\chi_1$) and fifth order susceptibility ($\chi_5$) for the same compounds respectively. (c) Temperature (in logarithmic scale) response of normalized resistivity ($\rho/\rho_{50K}$) for the series of compounds $Ce_{1-x}Y_xNiGe_2$ (x = 0.0, 0.1, 0.2, and 0.4) at 0 T. The data for Y-0.0 compound is from Ref [17]. Inset: Magnified plot of same figure in temperature range 2.8-5 K.

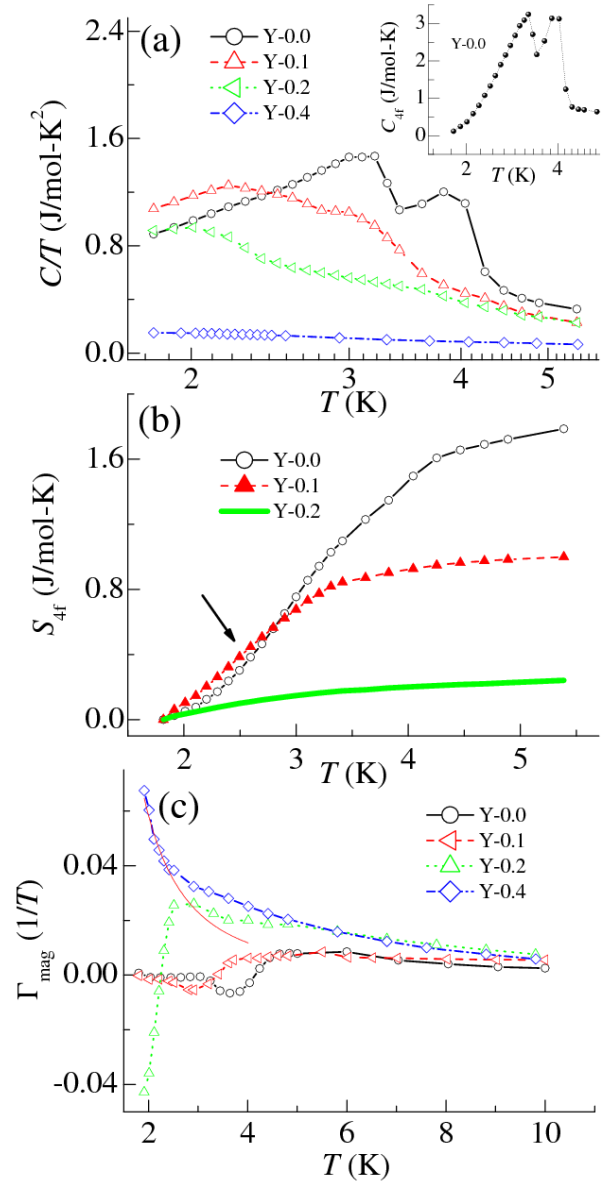

**Figure 6**: (a) Temperature (in logarithmic scale) response of heat capacity divided by temperature ($C/T$) for the series of compounds $Ce_{1-x}Y_xNiGe_2$ (x = 0.0, 0.1, 0.2, and 0.4) at 0 T. Inset of Fig. shows temperature (in logarithmic scale) response of $C_{4f}$ (data is taken from Ref (17)) for Y-0.0 compound. (b) Temperature response of magnetic entropy for all the compounds $Ce_{1-x}Y_xNiGe_2$ (x = 0.0, 0.1, 0.2). (c) Magnetic Gruneisen parameter ($\Gamma_{mag}$) versus temperature plots at 0.1 T for the series of compounds $Ce_{1-x}Y_xNiGe_2$ (x = 0.0, 0.1, 0.2, and 0.4). Red line through the data points represents the fitting of equation $\Gamma_{mag} \sim T^\epsilon$.